\documentclass[12pt,preprint]{aastex}
\newcommand{\pref}{\protect\ref}
\newcommand{\solrad}{\ifmmode{R}_{\rm S}\else${R}_{\rm S}$\fi}
\newcommand{\solmas}{\ifmmode{M}_{\rm S}\else${M}_{\rm S}$\fi}
\newcommand{\tintu}{$\,$ergs$\,$cm$^{-2}\,$s$^{-1}\,$sr$^{-1}$}

\newcommand{\ctn}{\ifmmode\kappa\else$\kappa$\fi}

\newcommand{\flxu}{$\,$ergs$\,$cm$^{-2}\,$s$^{-1}$}
\newcommand{\velu}{$\,$km$\,$s$^{-1}$}
\newcommand{\dynu}{$\,$dyn$\,$cm$^{-2}$}
\newcommand{\term}[2]{\mbox{$\,^{#1}{\rm #2}$}}
\def\term#1 #2/{\mbox{$\,^{#1}{\rm #2}$}}

\def\aspcs{{ASP Conf.\ Ser.}}

\renewcommand{\vec}[1]{{\bf #1}}
\newcommand{\cross}{\times}

\newcommand{\jcb}{\ifmmode\vec{j}\cross\vec{B}\else$\vec{j}\cross\vec{B}$ \fi}

\newcommand{\itime}{\ifmmode\tau_{1\kappa}\else$\tau_{1\kappa}$ \fi} 
\newcommand{\rtime}{\ifmmode\tau_{\kappa1}\else$\tau_{\kappa1}$ \fi} 
\newcommand{\etime}{\ifmmode\tau_{12}\else$\tau_{12}$ \fi} 
\newcommand{\mtime}{\ifmmode\tau_{np}\else$\tau_{np}$ \fi} 
\newcommand{\cttime}{\ifmmode\tau_{CT}\else$\tau_{CT}$ \fi} 
\newcommand{\nntime}{\ifmmode\tau_{nn}\else$\tau_{nn}$ \fi} 
\newcommand{\mfp}{\ifmmode\lambda\else$\lambda$\fi} 
\newcommand{\vmean}{\ifmmode\overline v\else$\overline v$\fi} 
\newcommand{\thick}{\ifmmode\Delta\else$\Delta$\fi}

\newcommand\figone{
\begin{figure}[!ht] 
\epsscale{0.9}
\plotone{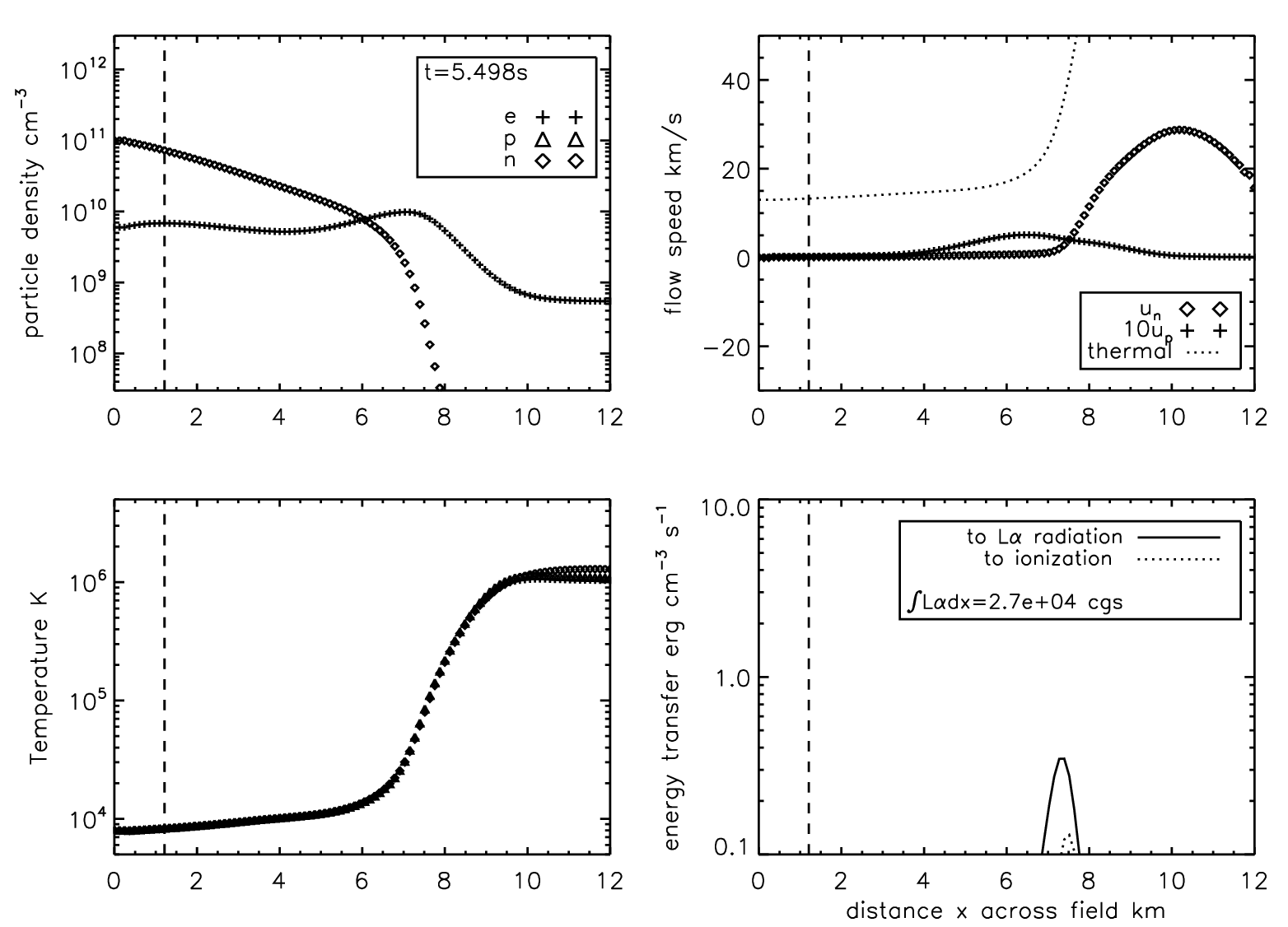}  
\caption{\label{fig:snapshot} Conditions a few seconds after hydrogen
  is allowed to diffuse across field lines into coronal plasma.   The
  abscissa is distance $x$ across the field lines, the initial cool flux
  tube extends from zero to the dashed line.  
}
\end{figure}
}

\newcommand\tabzero{
\protect\begin{deluxetable}{llll}
\tablecaption{Tube properties\label{tab:tube}}
\tablehead{Quantity & Inside & Outside \\ & (cool)  & (hot) } 
\startdata
radius $r_c$ cm & $5\times10^7$ & \\ 
length $l_c$ cm & $5\times10^8$ & $\ge 10l_c$\\ 
$T$  K      & $10^4$ & $10^6$ \\
$n_H$ cm$^{-3}$ & $8\times 10^{10}$ & $\approx0$ \\
$n_p$  cm$^{-3}$ & $n_H/40$ & $4\times10^8$ \\
$n_e$ cm$^{-3}$ & $n_H/40$ & $4\times10^8$ \\
$p$  \dynu{}   & 0.11 & 0.11 \\
Magnetic field strength $B$ G & 10 & 10\\
$B^2/8\pi$ \dynu{}    & 3.8 & 3.8 \\
\enddata 
\end{deluxetable}
}

\newcommand\tabone{
\protect\begin{deluxetable}{llllll}
\tablecaption{Plasma conditions \label{tab:regimes}}
\tablehead{&Quantity & Units & & scaling & notes} 
\startdata
\multispan{4}{Initial corona\hfill}\\
&$                           T_h$&                K& $  10^{6} $& \\
&$                           n_h$&             cm$^{-3}$& $  8.0\times10^8 $&\\
&$                       n_p,n_e$&             cm$^{-3}$& $  4.0\times10^8 $&\\
&$                             p$&             cm$^{-3}$& $  1.1\times10^{-1} $&\\
&$                             B$&                G& $  10 $&\\
&$                          \beta$&                 & $  2.8\times10^{-2} $&\\
&$                        \omega_p$&              s$^{-1}$& $  9.6\times10^4 $&\\
&$                         r_{gyro}$&               km& $  1.5\times10^{-3} $&\\
&$                 \tau_{pp}$&                s& $  1.6 $& $n_p^{-1} T^{+3/2}$\\
&$          \omega_p \tau_{pp}$&                 & $  1.5\times10^5 $&\\
&$                 \tau_{ee}$&                s& $  5.0\times10^{-2} $& $n_e^{-1} T^{3/2}$\\
\multispan{4}{chromospheric tube\hfill}\\
&$                           T_c$&                K& $  8.0\times10^3 $& \\
&$                         \vmean$&           km s$^{-1}$& $  13 $& $T^{1/2}$\\
&$                           n_c$&             cm$^{-3}$& $  10^{11} $&\\
&$                 \tau_{nn} $&                s& $  1.4\times10^{-2} $& $n_n^{-1} T^{-1/2}$\\
\multispan{4}{hot protons impacting hydrogen atoms\hfill}\\
&$                  \tau_{pn} (CT)$&                s& $
1.0\times10^{-2} $& $n_p^{-1} T^{-1/2}$ &{``$CT$'' = charge transfer}
\\
&                    H atom mfp&               km& $  6.5\times10^{-2} $& $n_p^{-1}$\\
\multispan{4}{cool hydrogen atoms impacting protons\hfill}\\
&$                  \tau_{np} (CT)$&                s& $  8.0\times10^{-5} $& $n_n^{-1} T^{-1/2}$\\
&                    proton mfp&               km& $  5.8\times10^{-3} $& $n_n^{-1}$\\
&$          \omega_p \tau_{np}$&                 & $  7.7 $&\\
\multispan{4}{hot electrons impacting H atoms\hfill}\\
&$                       \tau_{12}$&                s& $
9.5\times10^{-2} $& $n_e^{-1} T_e^{-1/2} e^{10.2e/kT_e}$& excitation
of $n=2$ level\\
&$                       \tau_{1k}$&                s& $
8.2\times10^{-2} $& $n_e^{-1} T_e^{-1/2} e^{13.6e/kT_e}$& ionization\\
&$                       \tau_{k1}$&                s& $
4.0\times10^5 $& $n_e^{-1} T_e^{+1/2}$& radiative recombination\\
\enddata 
\tablecomments{$\tau_{ab}$ refers to the time taken for a particle of
  type $b$ to be impacted by a sea of particles of type $a$, except
  where noted.}
\end{deluxetable}

}

\slugcomment{}

\begin{document}

\title{\large An explanation of the solar transition region}

\author{Philip Judge}
\affil{High Altitude Observatory,
National Center for Atmospheric Research\altaffilmark{1},
P.O. Box 3000, Boulder CO~80307-3000, USA\\ \vbox{}}

\altaffiltext{1}{The National Center for Atmospheric Research is
sponsored by the National Science Foundation}

\begin{abstract}

Prompted by high resolution observations, 
I propose an explanation for the 40+ year old problem of structure and
energy
balance in the solar transition region. The ingredients are simply
cross-field diffusion of neutral atoms from cool threads
extending into the corona, and the subsequent 
excitation, radiation 
and ionization of these atoms via electron impact.  
The
processes  occur whenever chromospheric plasma is adjacent to
coronal plasma, and are efficient even when ion gyro-frequencies exceed collision
frequencies. 
Cool threads - fibrils and spicules perhaps -  grow slowly in thickness as a
neutral, ionizing front expands across the magnetic field
into coronal plasma.  
Radiative intensities estimated for H L$\alpha$ are 
within an order of magnitude of those observed, with no ad-hoc 
parameters - only thermal parameters and geometric considerations are
needed.  I speculate that the subsequent dynamics of the diffused material
might also explain observed properties of trace elements.
\end{abstract}


\section{Introduction}
\label{sec:introduction}

The upper transition region (henceforth, ``TR'') - plasma with
electron temperatures in the range $2\times 10^5 \la T_e \la 10^6$ K,
is adequately described by field-aligned thermal conduction down from
the corona.  The lower TR ($10^4$ K $< T_e < 2\times 10^5$ K) however,
is not so easily understood \citep{Gabriel1976,Jordan1980b}.  Models
dominated by field-aligned heat conduction produce too little emission
from the lower TR by orders of magnitude, a problem already evident in
work by \citet{Athay1966}. Neither could such models radiate away the
downward directed conductive flux of $F_{cond}\sim 10^6$ \flxu{}
\citep[e.g.][]{Jordan1980b,Athay1981}.
\nocite{Fontenla+Avrett+Loeser2002} Fontenla {\em
et al.} (2002 and earlier papers in the series), henceforth ``FAL''
showed that energy balance can be achieved through field-aligned (1D)
diffusion of neutral hydrogen and helium atoms.  The neutral atoms
diffuse into hot regions, radiate away much of the coronal energy, and
can reproduce the H and He line intensities.

The problem might be considered by some as solved, in principle. 
But there exists the serious and nagging problem of the peculiar spatial relationship
between the observed corona, TR and chromosphere \citep{Feldman1983}.  
Feldman and colleagues have since analyzed many observations, concluding that 
the lower TR is thermally disconnected from
the corona \citep[e.g.][and references
therein]{Feldman+Dammasch+Wilhelm2001}.  
Yet 
\citet{Fontenla+Avrett+Loeser1990} declared that ``The above
[i.e. their] scenario explains why (as noted by Feldman 1983) the
\nocite{Feldman1983}
structure of the transition region is not clearly related to the
structures in the corona''.  
That the debate still rages
is evidenced by advocates for  ``cool
loop'' models in which lower TR radiation originates from loops
never reaching coronal temperatures and having negligible conduction 
\citep[][and references therein, henceforth ``PGV'']{Patsourakos+Gouttebroze+Vourlidas2007}.  
Here I propose a different scenario, prompted by new data and analyses which
show that neither cool loops nor field-aligned processes adequately 
describe the L$\alpha$ chromospheric network. I speculate that 
other TR lines might also be accounted for. 

\section{A new scenario}

L$\alpha$ network emission, at $0\farcs3$
resolution appears mostly as 
threads  of relatively uniform
intensity, of 5-10 Mm length and 1Mm diameter (PGV).  PGV argued
that ``the different appearance the TR has in the quiet Sun
[i.e. network] is suggesting that the bulk of its emission comes from
structures other than the footpoints of hot loops''.  
Convolved L$\alpha$ images from PGV
appear to correspond to those seen 
in many other TR lines at lower resolution 
\citep[e.g.][]{Curdt+others2001}. 
\citet{Judge+Centeno-Elliott2008} showed, using magnetic field
measurements from Kitt Peak, that much of the network L$\alpha$ emission originates
in long spicule-like structures lying along the lowest few Mm of
magnetic field lines extending into the corona, but that plage
emission may correspond to the thin footpoints as suspected by
PGV and modeled by FAL.  Even in plages, on
sub-arcsecond scales, field-aligned threads of cool plasma (fibrils,
spicules),  extend into the low corona forming non-planar
thermal interfaces between hot and cool plasma
\citep{Berger+others1999}.  Prompted by these data, I examine the diffusion of
neutral particles into the corona, {\em across} magnetic fields
\citep[following a suggestion by][]{Pietarila+Judge2004}.  

Consider a straight cylinder of cool, partially ionized material embedded
in a hot corona, of
radius $r_c$. Length $l_c \gg r_c$ of the 
tube contains cool plasma 
in contact with the
hot corona.  The magnetic tube is of length $L \gg l_c$,
mostly containing coronal plasma.
Tube parameters are given in Table \pref{tab:tube}. 
Note that the neutral density greatly
exceeds other densities. 
The chosen geometry
is typical of values found by
PGV, and thermal parameters
are typical of the quiet Sun\footnote{I do not adopt the higher temperatures of cool loops
  used by PGV, because here the corona and dynamics supplies all the 
  energy for L$\alpha$ emission.}.
The plasma is assumed to be
in a low plasma-$\beta$ regime. 

\subsection{Initial diffusion, relaxation, radiation}
\label{subsec:init}
Imagine an injection of dense neutral material into the tube footpoint by
some 
chromospheric process.
The tube surface acts as a
semi-permeable membrane.  Neutral particles travel freely between
collisions, but ions gyrate about magnetic field lines with gyro radii 
orders of magnitude smaller than mean free paths 
(``mfp''s, Table~\pref{tab:regimes}).  Ions and electrons
are essentially frozen to field
lines, but neutrals can diffuse across field lines
almost as efficiently as along them, 
and find themselves impacted by hot electrons and protons.   

Table \pref{tab:regimes}  lists time scales for kinetic processes for 
a ``cool'' hydrogen atom embedded in a hot corona of $T=10^6$K,
using data from
\citet{Allen1973},
\citet[][henceforth HLH]{Hansteen+Leer+Holzer1997}, and
\citet{Gilbert+Hansteen+Holzer2002}.  A hydrogen atom
crossing the boundary encounters other diffusing hydrogen atoms and hot 
protons and electrons.  
Statistically, the first interaction is a
collision with a coronal proton, involving the exchange of energy 
and ($\sim50$\% of the time) 
an electron (charge transfer, ``CT'').  Charge transfer yields an exchange of
momentum (180$^\circ$ change in direction) but little exchange of
energy \citep[e.g.][]{Osterbrock1961}.
The kinetic energy exchanged 
is $\sim \frac{3}{2}kT_h$, shared between
them after two such collisions 
(I use subscript ``$h$'' to
denote hot and ``$c$'' cool plasma). 
The CT cross section is roughly independent of
energy, so the ``warm''  neutral atom  has a $\approx 1- e^{-1}=0.63$
probability of staying within the hot plasma. Assuming that it does so,
after
$\itime\approx 8 \cttime$ s it becomes ionized by impact with a hot
electron. Once free, the electron will not readily recombine with a
proton (time scale $\rtime \sim 3\times10^{5}$ s).  At $T_e=T_h=10^6$K, the
time $\etime$ needed for electron impact excitation of the $n=2$
levels and (rapid) emission of a L$\alpha$ photon
is comparable to $\itime$.  Thus, of the ionized neutral atoms,
$\approx 50\%$ will have emitted a L$\alpha$ photon, the energy
supplied by coronal electrons and protons. Because there are
relatively few hot particles, their thermal energy limits the
number of neutral atom ionizations and excitations.

From kinetic theory, the flux density of neutral hydrogen
atoms {\em initially} crossing the boundary into the corona is 
$
\frac{1}{4} n_c \vmean_c \sim 2\times10^{16}$ particles~s$^{-1}$~cm$^{-2}$,
where $\vmean{}=
\sqrt{8kT_c/\pi m}$.  The kinetic energy per ``hot'' proton is
$\frac{3}{2}kT_h$ which is shared roughly equally after two CT
collisions with a neutral, producing a population of $n_h
\ll n_c$ ``warm'' neutrals with $T\sim T_h/2$.  After a few, say $m$
more collisions (time $m\,\cttime$ later), all the proton energies are the
larger of $\sim \frac{3}{2}kT_h/2^m$ and $\frac{3}{2}kT_c$, and
$m n_h$ of the $n_c$ neutrals have suffered a proton impact. (A time
of $n_c/n_h =100$ times \cttime{} $\approx 1$s is required before all neutrals have
been impacted). 
The  warm
neutrals  relax via collisions with
the cool neutrals. 

The initial electron evolution is largely determined by inelastic
collisions with hydrogen:
each hot electron typically has sufficient energy to excite and ionize
5 neutral hydrogen atoms, which takes 
$\sim7\itime \sim 0.6$s.  (Electron-electron collision times are $\la
0.05$ s). 
The electrons lose energy ${\varepsilon}=5n_h(I+E)e$ per unit volume 
at the rate 
\begin{equation} \label{eqn:hlossrate}
\frac{\varepsilon}{t}  \ga
\frac{5n_h(I+E)e}{7\itime}\approx 0.13 {\rm \ \  erg~cm^{-3}~s^{-1}},
\end{equation}
leading to a cooling time of $\la 0.4$ s. (A lower limit applies because
tails of 
the Maxwellian distributions can increase excitation/ionization
rates).  Of this energy a fraction 
$\frac{E}{E+I} = \frac{3}{7}$ is emitted in L$\alpha$. 
The radial flux density of L$\alpha$ radiation from this neutral ``sheath'' is 
\begin{equation} \label{eqn:laphaf}
f \ga \frac{3}{7}\frac{\varepsilon}{t} \thick \equiv
\frac{3}{7}{\varepsilon} v_c^{diff} {\rm \ \  erg~cm^{-2}~s^{-1}},
\end{equation}
\noindent where $\thick$ is the sheath thickness
at time $t$ ($60 \cttime$), and $v_c^{diff} = \thick/t$ is the
diffusion speed.  For a random
walk, 
$\thick_c \approx \frac{1}{3} \sqrt{60} \cttime \vmean_c \approx
3.3\times 10^4$ cm, for
warmed neutrals $\thick_w \approx 1.8\times10^5$ cm ($v_c^{diff}=0.57$
and $v_w^{diff}=3.2$ km/s respectively;  the factor $\frac{1}{3}$  accounts for 
the random direction of the ``walk'').  As a rough
estimate, I take 
$v^{diff} \approx 3v^{diff}_c$ km/s:
\begin{equation} \label{eqn:laphaf}
f \approx  \frac{3}{7}{\varepsilon} 3 v_c^{diff} \approx
5.6\times10^3  {\rm \ \  erg~cm^{-2}~s^{-1}}
\end{equation}
Thus, $f$ 
is initially just a fixed fraction of 
the local coronal 
energy density multiplied by the diffusion speed. 
The specific intensity $I$ equals $f/\pi$ when the line is optically
thick and all the radiation scatters away from the solar surface.  (Photon
mfps for L$\alpha$ in the sheath are just $10^2$ cm).  This estimate
of $I$ is a factor of 100-300 below measured values of $(1.8-5.6)\times10^5$
\tintu{}  in active network threads
(PGV) and 30 below
average network intensities \citep{Vernazza+Reeves1978}.  
But, as will be made clear below,  
this is an under-estimate.  
Similar estimates of intensities for H L$\beta$ and the 584\AA{}
line of He~I, relative to L$\alpha$, 
are quite reasonable, recognizing that L$\beta$ is
optically thick across the sheath.  

\subsection{A multi fluid calculation}
\label{subsec:ddt}

To examine the evolution at later times, 
multi-fluid equations for conservation of mass, momentum and energy
were solved as functions of time and distance
$x$ across the field lines following \citet{Schunk1977} and HLH. 
Just electrons, protons and
neutral hydrogen atoms were treated.
Cartesian geometry is used because the diffusion region is much
thinner than the tube.
I assume that electrons are strongly tied to protons, so that their
densities and fluid velocities are equal ($n_e=n_p$, $u_e=u_p$: 
charge and electrical currents are neglected). 
The conservation equations used for 
mass, momentum and energy density for the fluid of 
species $s$ are
\newcommand\ddt[1]{\frac{\partial #1}{\partial t}}
\begin{equation} \label{eqn:nconserv}
\ddt{n_s} +  \frac{ \partial}{\partial x}  {\left \{ n_s u_s +
  d^n_s\right \} } = \frac{\delta n_s}{\delta t}, 
\end{equation}
\begin{equation} \label{eqn:momconserv}
m_s\ddt{n_s u_s} +  \frac{ \partial}{\partial x}  {\left \{ m_s n_s u_s^2 +
  p_s+ d^M_s\right \} } +F = \frac{\delta M_s}{\delta t}, 
\end{equation}
\begin{equation} \label{eqn:econserv}
\noindent \ddt{E_s } +  \frac{ \partial}{\partial x}  {\left \{ u ( E_s +
  p_s)+ d^E_s \right \} } = \frac{\delta E_s}{\delta t} +Q-L.
\end{equation}
No conservation equation is used for the heat flux since here it is
treated as $d_s^E$ using the mfp approximation. 
Above, $F$ is a body force term (gravity, Lorentz force for example), 
$E_s=\frac{3}{2}n_skT_s+ \frac{1}{2}
m_s n_s u_s^2$, $p_s=n_skT_s$, and the $\frac{\delta\  }{\delta t}$ are
non-linear
collisional terms.    
$Q$ and $L$
are the 
energy gains and losses respectively, where 
I adopt 
$Q=1.67\times10^{-25} n_en_p e^{-T_H/8000}$ erg~cm$^{3}$~s$^{-1}$ to
maintain a chromosphere against losses $L$
(HLH),
and $L$ includes latent heat and L$\alpha$ radiative
losses computed explicitly from the collisional terms.  

The diffusion
terms $d_s$ (not included by
HLH, except for the heat flux) require care especially for the
dynamics of the proton fluid. For individual protons and electrons, 
the momentum equations are dominated by the 
Lorentz force.
Their cross-field motion on timescales
short compared with collision times is circular with frequency
$\omega_s=e_sB/m_s$. On longer time scales 
the summed (fluid parcel) momenta can change only after a collision.  The net
effect of the Lorentz force is thus to limit the cross-field displacement
of charged particles to a single 
gyro radius
$r_s=\vmean_s/\omega_s$ in collision time $\tau_s$ instead of the collisional mean free path
$\lambda_s=\vmean_s\tau_s$.   Thus, a simple recipe for calculating 
cross-field transport via the fluid equations 
is to set both $F$ and $\frac{\partial p_p}{\partial x}$ 
terms to zero in the proton momentum equation, and modify the $d_p$
terms to account for the reduced displacements.  Field-free 
diffusion is described by  equations (4.41),
(4.46) and (4.52) of \citet{Gombosi1994}:
\begin{equation} \label{eqn:fluxes}
d^n_s = -\frac{1}{3} \lambda_s \frac{\partial}{\partial x} \left \{ n_s
\overline{v}_s \right \}, \ \ \ \ \ 
d^M_s = -\frac{1}{3} \lambda_s \frac{\partial}{\partial x} \left
\{ m_s n_s u_s
\overline{v}_s \right \}, \ \ \ \ \ 
d^E_s = -\frac{\pi}{12} \lambda_s \frac{\partial}{\partial x}
\left \{ n_s m_s
\overline{v}^3_s \right \},
\end{equation}
\noindent For charged particles, $\lambda_s$ must be replaced by 
$\lambda^\ast_s= \lambda_s/(1+\omega_s\tau_s)^2$ (following the above
argument, see
\citealp[][eqs. 4.37, 4.40]{Braginskii1965}). 
Note that, written in terms of $T_s$, $d_s^E$ yields
the widely used ``Spitzer'' thermal conductivity parallel to the
field, and the ion-dominated conductivity perpendicular to the field.
The net effect for $\omega_s\tau_s \gg 1$ is that only the neutral fluid 
diffuses efficiently across the field- the charged fluid evolves mostly via the
collisional coupling to the neutrals (via $\delta M_s/\delta t$), 
and to a lesser degree to the small $d_s$ terms.

The 
variables $(n_s, u_s, E_s)$ for electrons, protons and neutral H
atoms, functions of $(t;x)$, 
were initialized according to table~\pref{tab:tube}.  Only 7 variables
were solved since it is assumed that $n_e=n_p$ and $u_e=u_p$. 
The equations were 
integrated in time using MacCormack's
method to include the collisional terms 
\citep{Griffiths+Higham1999}.  For the first three
points near $x=0$, the variables were held fixed to their initial
values, maintaining the same chromospheric conditions there.
Figure~\pref{fig:snapshot} shows conditions several seconds after the
beginning of the diffusion process.  Pressure gradients
drive neutrals into the corona against friction forces, thus the 
diffusion speed, measured by tracking the steep temperature rise, 
is $\approx0.8$ \velu, far below the thermal speed. 
The computed 
flux density of L$\alpha$ is $\approx 5\times10^4$ \flxu{}, and 
is roughly constant in time.  It is some 10$\times$ higher than the simple kinetic
result above, because of the nonlinear dynamics: (1) the densities become higher in the corona, 
(2) flow energy is converted to heat, (3) the L$\alpha$
losses/latent heat ratio is higher (the photons are created at
electron temperatures lower than the initial coronal temperature).  
$I$ is computed to be just a factor of 10-30 below 
observed active network thread intensities, and 3 below 
average network intensities.

A calculation with twice the coronal density, more
appropriate for active network, yields smaller diffusion speeds and
L$\alpha$ fluxes which are just 1.7 times higher. EUV/X-ray
coronal intensities  scale with (density)$^2$, and so 
would be
a factor of four brighter.  This non-linear relationship is an
important property of the calculations.

\section{Discussion, speculations}

Based upon observations of spicules and other fine, thread-like
structures on the solar disk, it is clear that non-planar thermal interfaces
exist at the base of the corona, and that the morphology of the TR emission 
from such interfaces cannot be explained by field-aligned particle transport at
the base of coronal loops, in contrast to the  claims by 
\citet{Fontenla+Avrett+Loeser1990}.  
The picture proposed here uses unspecified mechanisms in the chromosphere
to maintain a reservoir of cool mostly neutral plasma
directly adjacent to hot coronal plasma.  The cylindrical
geometry, inspired by observations, presents a large surface area (per
unit volume) of contact between cool and hot plasma.  The chromosphere
supplies mass via neutral diffusion across the surface to a thermal
boundary layer, and the corona supplies energy to the neutral
particles. The originally neutral particles drain energy from the corona
by latent heat of ionization and by inelastic collisions leading to
strong L$\alpha$ emission. The
diffusing layer propagates outwards, emitting radiation like the
boundary of a wild fire\footnote{Secchi in 1877 described the
chromosphere as a ``burning prairie'', but in a different sense.},
into the corona until either the supply of neutral mass or coronal
energy dries up.  The present 
proposal is related to models invoking 
cross-field heat conduction \citep{Rabin+Moore1984,Athay1990}.  This 
effect is included here (via $d_s^E$), but it is far less efficient 
at moving heat to cool plasma than  diffusion is at moving neutral
atoms to the coronal heat.

The calculations presented here fall short of accounting for the large
radiative flux of L$\alpha$, by factors of $\approx 10$.  However,
the calculations miss important additional sources of energy in the corona:
thermal and gravitational potential energy.  The cool threads extend
only a few Mm into the corona, and form just the lower parts
of a much larger coronal structure.  The diffused cool material is
thus subject to parallel transport (heat conduction, diffusion) which
will transfer heat from the overlying coronal plasma to the diffused
material.  Spicules formed by ejection from the chromosphere will have
their entire length exposed to this energy flux, because the lowest
parts of the spicules diffuse first into the corona- the diffusion
fronts are not exactly parallel to field lines.  Coronal plasma along
connected field lines contains $L \frac{3}{2} nkT$ erg~cm$^{-2}$,
where $L$ is the pressure scale height ($\sim 50$Mm) or loop
length. Since $L \gg l_c$ the energy available for L$\alpha$ radiation
would be $L/l_c \ge 10\times$ larger than computed above, more if the tubes expand with
height.  I speculate that cross-field diffusion and subsequent
parallel conduction {\em might bring theoretical and observed
intensities values into agreement.}  The time needed to conduct this
energy must lie between the electron sound speed $c_e$ as $L/c_e \sim
13$ s, and $\sim 10^3$ s, an upper limit obtained from the thermal
energy divided by the conductive flux for a uniform temperature
gradient.  Gravitational potential energy might contribute to the
heating and dynamics of the sheath as the diffused material cools
the corona and adds mass, such that vertical pressure balance no
longer is expected.  It may be that larger red-shifts would be
expected where magnetic fields are more vertical, i.e. directly over
the magnetic network.  This expectation is not in disagreement with results
found by \citet{McIntosh+others2007}.  However,
little more can be said without solving the 2D multi-fluid
conservation equations including parallel heat conduction and
cross-field diffusion, beyond the scope of this letter.  Such
calculations will also show if the emission lines of trace species
(ions of carbon, oxygen etc. in the TR) can be explained. 

Cool threads are observed in different coronal environments (PGV)-
their intensities appear to vary relatively little compared with the
embedding coronal intensities. This fact is part of Feldman's (1983)
\nocite{Feldman1983} claim that TR emission is energetically
disconnected from the corona.   The calculations presented here indeed
produce
a non-linear relationship between L$\alpha$ and coronal
brightness. The L$\alpha$ intensities scale with the local coronal
energy density and with the diffusion speed. But the EUV and X-ray
radiation emitted by the corona itself vary with density$^2$ and
peaked functions of temperature along lines of sight different from
the direction of field lines into the sheath. The scenario might
therefore explain most of the observed puzzling facets noted by
Feldman and colleagues, yet still maintain a strong energetic link
between the corona and TR, and thereby resolve a long-standing debate
(see the different perspectives of
\citealp{Feldman+Dammasch+Wilhelm2001} and
\citealp{Wikstol+Judge+Hansteen1998}, for example).

To see if the scenario survives scrutiny,
more observations of
chromospheric fine structure and its relation with the corona and TR
would be as important as numerical modeling work.

\noindent I am grateful to Tom Holzer, Scott McIntosh and the referee
for comments. 

\def\aspcs{{ASP Conf.\ Ser.}}

\tabzero
\tabone
\figone

\begin{thebibliography}{}

\bibitem[\protect\astroncite{Allen}{1973}]{Allen1973}
Allen, C.~W.: 1973,
\newblock {\em Astrophysical Quantities\/},
\newblock Athlone Press, Univ.\ London

\bibitem[\protect\astroncite{Athay}{1966}]{Athay1966}
Athay, R.~G.: 1966,
\newblock {\em Astrophys.\ J.\/} {\bf 145}, 784

\bibitem[\protect\astroncite{Athay}{1981}]{Athay1981}
Athay, R.~G.: 1981,
\newblock {\em Astrophys.\ J.\/} {\bf 249}, 340

\bibitem[\protect\astroncite{Athay}{1990}]{Athay1990}
Athay, R.~G.: 1990,
\newblock {\em Astrophys.\ J.\/} {\bf 362}, 364

\bibitem[\protect\astroncite{{Berger} {\em et~al.}}{1999}]{Berger+others1999}
{Berger}, T.~E., {De Pontieu}, B., {Schrijver}, C.~J., and {Title}, A.~M.:
  1999,
\newblock {\em Astrophys.\ J.\ Lett.\/} {\bf 519}, L97

\bibitem[\protect\astroncite{Braginskii}{1965}]{Braginskii1965}
Braginskii, S.~I.: 1965,
\newblock {\em Reviews of Plasma Physics.\/} {\bf 1}, 205

\bibitem[\protect\astroncite{{Curdt} {\em et~al.}}{2001}]{Curdt+others2001}
{Curdt}, W., {Brekke}, P., {Feldman}, U., {Wilhelm}, K., {Dwivedi}, B.~N.,
  {Sch{\"u}hle}, U., and {Lemaire}, P.: 2001,
\newblock {\em Astron.\ Astrophys.\/} {\bf 375}, 591

\bibitem[\protect\astroncite{Feldman}{1983}]{Feldman1983}
Feldman, U.: 1983,
\newblock {\em Astrophys.\ J.\/} {\bf 275}, 367

\bibitem[\protect\astroncite{{Feldman} {\em
  et~al.}}{2001}]{Feldman+Dammasch+Wilhelm2001}
{Feldman}, U., {Dammasch}, I.~E., and {Wilhelm}, K.: 2001,
\newblock {\em Astrophys.\ J.\/} {\bf 558}, 423

\bibitem[\protect\astroncite{Fontenla {\em
  et~al.}}{1990}]{Fontenla+Avrett+Loeser1990}
Fontenla, J.~M., Avrett, E.~H., and Loeser, R.: 1990,
\newblock {\em Astrophys.\ J.\/} {\bf 355}, 700

\bibitem[\protect\astroncite{{Fontenla} {\em
  et~al.}}{2002}]{Fontenla+Avrett+Loeser2002}
{Fontenla}, J.~M., {Avrett}, E.~H., and {Loeser}, R.: 2002,
\newblock {\em Astrophys.\ J.\/} {\bf 572}, 636 (FAL)

\bibitem[\protect\astroncite{Gabriel}{1976}]{Gabriel1976}
Gabriel, A.: 1976,
\newblock {\em Phil Trans. Royal Soc. Lond.\/} {\bf 281}, 339

\bibitem[\protect\astroncite{{Gilbert} {\em
  et~al.}}{2002}]{Gilbert+Hansteen+Holzer2002}
{Gilbert}, H.~R., {Hansteen}, V.~H., and {Holzer}, T.~E.: 2002,
\newblock {\em Astrophys.\ J.\/} {\bf 577}, 464 

\bibitem[\protect\astroncite{Gombosi}{1994}]{Gombosi1994}
Gombosi, T.~I.: 1994,
\newblock {\em Gaskinetic Theory\/},
\newblock Cambridge University Press, Cambridge, England

\bibitem[\protect\astroncite{Griffiths and Higham}{1999}]{Griffiths+Higham1999}
Griffiths, D. and Higham, D.: 1999,
\newblock {\em MacCormack's method for advection-reaction equations\/},
\newblock Technical report, Department of Mathematics, University of
  Strathclyde

\bibitem[\protect\astroncite{Hansteen {\em
  et~al.}}{1997}]{Hansteen+Leer+Holzer1997}
Hansteen, V., Leer, E., and Holzer, T.: 1997,
\newblock {\em Astrophys.\ J.\/} {\bf 482}, 498 (HLH)

\bibitem[\protect\astroncite{Jordan}{1980}]{Jordan1980b}
Jordan, C.: 1980,
\newblock {\em Astron.\ Astrophys.\/} {\bf 86}, 355

\bibitem[\protect\astroncite{Judge and
  Centeno}{2008}]{Judge+Centeno-Elliott2008}
Judge, P.~G. and Centeno, R.: 2008,
\newblock {\em Astrophys.\ J.\/} in press

\bibitem[\protect\astroncite{{McIntosh} {\em
  et~al.}}{2007}]{McIntosh+others2007}
{McIntosh}, S.~W., {Davey}, A.~R., {Hassler}, D.~M., {Armstrong}, J.~D.,
  {Curdt}, W., {Wilhelm}, K., and {Lin}, G.: 2007,
\newblock {\em Astrophys.\ J.\/} {\bf 654}, 650

\bibitem[\protect\astroncite{{Osterbrock}}{1961}]{Osterbrock1961}
{Osterbrock}, D.~E.: 1961,
\newblock {\em Astrophys.\ J.\/} {\bf 134}, 347

\bibitem[\protect\astroncite{{Patsourakos} {\em
  et~al.}}{2007}]{Patsourakos+Gouttebroze+Vourlidas2007}
{Patsourakos}, S., {Gouttebroze}, P., and {Vourlidas}, A.: 2007,
\newblock {\em Astrophys.\ J.\/} {\bf 664}, 1214 (PGV)

\bibitem[\protect\astroncite{Pietarila and Judge}{2004}]{Pietarila+Judge2004}
Pietarila, A. and Judge, P.~G.: 2004,
\newblock {\em Astrophys.\ J.\/} {\bf 606}, 1239

\bibitem[\protect\astroncite{Rabin and Moore}{1984}]{Rabin+Moore1984}
Rabin, D. and Moore, R.: 1984,
\newblock {\em Astrophys.\ J.\/} {\bf 285}, 359

\bibitem[\protect\astroncite{{Schunk}}{1977}]{Schunk1977}
{Schunk}, R.~W.: 1977,
\newblock {\em Reviews of Geophysics and Space Physics\/} {\bf 15}, 429

\bibitem[\protect\astroncite{Vernazza and Reeves}{1978}]{Vernazza+Reeves1978}
Vernazza, J.~E. and Reeves, E.~M.: 1978,
\newblock {\em Astrophys.\ J.\ Suppl.\ Ser.\/} {\bf 37}, 485

\bibitem[\protect\astroncite{Wikst{\o}l {\em
  et~al.}}{1998}]{Wikstol+Judge+Hansteen1998}
Wikst{\o}l, {\O}., Judge, P.~G., and Hansteen, V.: 1998,
\newblock {\em Astrophys.\ J.\/} {\bf 501}, 895

\end{thebibliography}
\end{document}